# Metasurface Integrated Monolayer Exciton Polariton

*Yueyang Chen[1]†, Shengnan Miao[4]†, Tianmeng Wang[4]†, Ding Zhong[2], Abhi Saxena[1], Colin Chow[2], James Whitehead[1], Xiaodong Xu[2,3], Su-Fei Shi[4,5], Arka Majumdar[1,2,\*]*

[1] *Electrical and Computer Engineering, University of Washington, Seattle, WA 98189, USA*

[2] *Department of Physics, University of Washington, Seattle, WA 98189, USA*

[3] *Materials Science and Engineering, University of Washington, Seattle, WA 98189, USA*

[4] *Department of Chemical and Biological Engineering, Rensselaer Polytechnic Institute, Troy, New York 12180, USA*

[5] *Department of Electrical, Computer, and Systems Engineering, Rensselaer Polytechnic Institute, Troy, New York 12180, USA*

*† These authors contributed equally to this work.*

*[\*]Corresponding Author: arka@uw.edu*

## Abstract

Monolayer transition metal dichalcogenides (TMDs) are the first truly two-dimensional (2D) semiconductor, providing an excellent platform to investigate light-matter interaction in the 2D limit. Apart from fundamental scientific exploration, this material system has attracted active research interest in the nanophotonic devices community for its unique optoelectronic properties. The inherently strong excitonic response in monolayer TMDs can be further enhanced by exploiting the temporal confinement of light in nanophotonic structures. Dielectric metasurfaces are one such two-dimensional nanophotonic structures, which have recently demonstrated strong potential to not only miniaturize existing optical components, but also to create completely new class of designer optics. Going beyond passive optical elements, researchers are now exploring active metasurfaces using emerging materials and the utility of metasurfaces to enhance the light-matter interaction. Here, we demonstrate a 2D exciton-polariton system by strongly coupling atomically thin tungsten diselenide ($WSe_2$) monolayer to a silicon nitride (SiN) metasurface. Via energy-momentum spectroscopy of the $WSe_2$-metasurface system, we observed the characteristic anti-crossing of the polariton dispersion both in the reflection and photoluminescence spectrum. A Rabi splitting of 18 meV was observed which matched well with our numerical simulation. The diffraction effects of the nano-patterned metasurface also resulted in a highly directional polariton emission. Finally, we showed that the Rabi splitting, the polariton dispersion and the far-field emission pattern could be tailored with subwavelength-scale engineering of the optical meta-atoms. Our platform thus opens the door for the future development of novel, exotic exciton-polariton devices by advanced meta-optical engineering.

## Introduction

Monolayer transition metal dichalcogenides (TMDs) have generated active research interest in recent years due to their strong light-matter interaction and unique optoelectronic properties[1]. Thanks to the quantum confinement in the atomic layer, exciton with large binding energy can form in monolayer TMD at room temperature, exhibiting strong excitonic absorption and photoluminescence[2]. The strong excitonic response could be further enhanced by coupling the TMD monolayer to an optical cavity[3]. In the weak coupling regime, low-threshold optically pumped nano-laser[4–7] and cavity-enhanced light-emitting diodes[8] have been demonstrated using TMD monolayer. In the strong coupling regime, TMD exciton-polaritons have also been observed at room temperature[9–11]. The hybrid light-matter quasi-particle, known as exciton-polaritons, inherit the low effective mass from their photonic component and high interaction nonlinearity form their excitonic component, making them a promising platform to study the physics of superfluids and Bose–Einstein condensate[12], with far-reaching impact on quantum simulation with interacting photons[13]. The observation of TMD exciton-polariton at room temperature also has potential applications in low-power nonlinear optics[14] and polariton laser[15]. Furthermore, the TMD exciton-polariton inherits the unique spin-valley physics from its exciton part[16], and the optical valley hall effect of the TMD exciton-polariton could be explored in this hybrid light-matter system[17].

So far, most TMD-based exciton-polariton devices are based on distributed Bragg reflector (DBR) cavity[9,18,19]. The fabrication process of the DBR-sandwiched TMD platform is non-trivial since the encapsulation of the upper DBR layers often degrade the optical property of the monolayer TMD[20]. Plasmonic cavities have also been explored[21], but it suffers from intrinsic absorption loss of metals. Another promising platform will be sub-wavelength patterned surfaces, also known as dielectric metasurfaces[22]. These metasurfaces can shape the optical wavefront using the sub-wavelength scatterers, also known as meta-atoms, and have recently been used to drastically miniaturize imaging and sensing devices[23–25], as well as to enhance light-matter interaction[26]. This nanopatterned, periodic photonic lattice supports a rich cluster of optical Bloch mode and can tightly confine the electromagnetic field[27,28]. Moreover, computational inverse design and dispersion engineering of the optical meta-atoms allow unprecedented nanophotonic engineering for the hybrid light-matter system[29–31]. While tremendous progress happened in building passive metasurfaces, recently researchers are exploring integrating new materials to create active metasurfaces[32,33]. In fact, layered material GaSe has been coupled to a silicon metasurface to demonstrate nonlinear frequency conversion[34]. Formation of Exciton-polaritons has also been demonstrated on a TMD-coupled to one-dimensional periodic structure[35]. While efforts have been made to couple TMD exciton to two-dimensional periodic structure, to date only the weak-coupling regime has been reported[36,37]. The planarized structure of metasurfaces favors the evanescent coupling to the two-dimensional layered material, which considerably simplifies the fabrication process since the pristine monolayer can be directly transferred on the metasurface with no further capping process required.

In this paper, we demonstrated exciton-polaritons in atomically thin tungsten diselenide ($WSe_2$) strongly coupled to the guided mode resonances (GMR) in a silicon nitride (SiN) metasurface. This unique resonance features of the GMR simultaneously achieve strong confinement of photons inside the metasurface and efficient coupling with the radiation continuum. By performing energy-momentum spectroscopy on the $WSe_2$-SiN metasurface sample, we experimentally observed the anti-crossing of the polariton dispersion both in the cavity reflection and photoluminescence. The exciton-polariton dispersion measured in the experiment is also reproduced by our numerical simulation. Moreover, we showed that the Rabi splitting, the polariton dispersion and the far-field emission pattern could be tailored by subwavelength-scale engineering of the optical meta-atoms. Our platform opens the door for the future development of novel exciton-polariton devices by advanced meta-optical engineering.

# Results

**Guided mode resonances in SiN metasurface**

Figure 1a shows the schematic of our platform. The metasurface is made of SiN meta-atoms with square lattice of holes. The whole structure sits on a silicon dioxide substrate. A $WSe_2$ monolayer could be transferred directly on top of the metasurface for evanescent coupling. Such periodic two-dimensional photonic lattice supports many optical Bloch modes propagating inside the slab. These modes can be classified into two classes[38]. The first is called in-plane guided modes, which are completely confined inside the slab without any coupling with the radiation continuum. In the photonic band-diagram, these modes exist outside the light-cone[39], and have been used in large scale photonic integrated circuits and optical sensing[40,41]. The second is called guided mode resonance (GMR), where the modes lie inside the light-cone in the photonic band diagram. The GMRs still have their electromagnetic power strongly confined inside the slab. However, unlike the in-plane guided mode, they couple with the radiation continuum: an effect also known as resonant-type Wood's anomaly[23]. The photon leakage channel allows straightforward excitation and probing of the modes from the free space without any extra photonic components like grating couplers. When a light beam shines on the metasurface, the interference between the slab mode and the GMR modes result in a Fano lineshape in the reflection spectrum. These Fano resonances have been used in the past for dispersion engineering[42], spectral filtering[43], and enhancing the light-matter interaction[34,44].

We simulated the reflection spectrum of the GMR in the SiN metasurface through Rigorous coupled-wave analysis (RCWA)[45,46]. The square lattice has a period of 459nm and a hole diameter of 120nm. The metasurface has a thickness of 130nm and it sits on a 1 µm thermal oxide layer grown on 500 µm silicon substrate. SiN has a refractive index of 2.0 and the silicon oxide has a refractive index of 1.45 in the RCWA simulation. The sub-wavelength thickness guarantees more electrical field on the meta-atom's surface leading to a larger light-matter interaction since the WSe$_2$ monolayer would be evanescently coupled to the GMR via the surface field. The left panel of the Figure 1b shows the angle-dependent reflection spectrum along the $k_x$ direction for p-polarized incident light. There are two different GMRs in the spectrum (M1 and M2). One has a linear dispersion (M1) and starts at higher energy and rapidly goes to lower energy when it comes to higher momentum. The other has a parabolic shape (M2): the mode starts at lower energy and goes to higher energy. The two modes come close to each other at $k_x = 0.6\ \mu m^{-1}$, and an anti-crossing behavior appears due to the coupling between two photonic modes. The electric field distribution of two modes (M1 and M2) at $k_x = 0$ is shown in Figure 1c. Two modes have different symmetry in terms of the field distribution, but they are both well confined inside the metasurface. Recently, such coupling between the GMRs has been explored to sculpt photonic bound state in continuum (BIC) state through dispersion engineering[47].

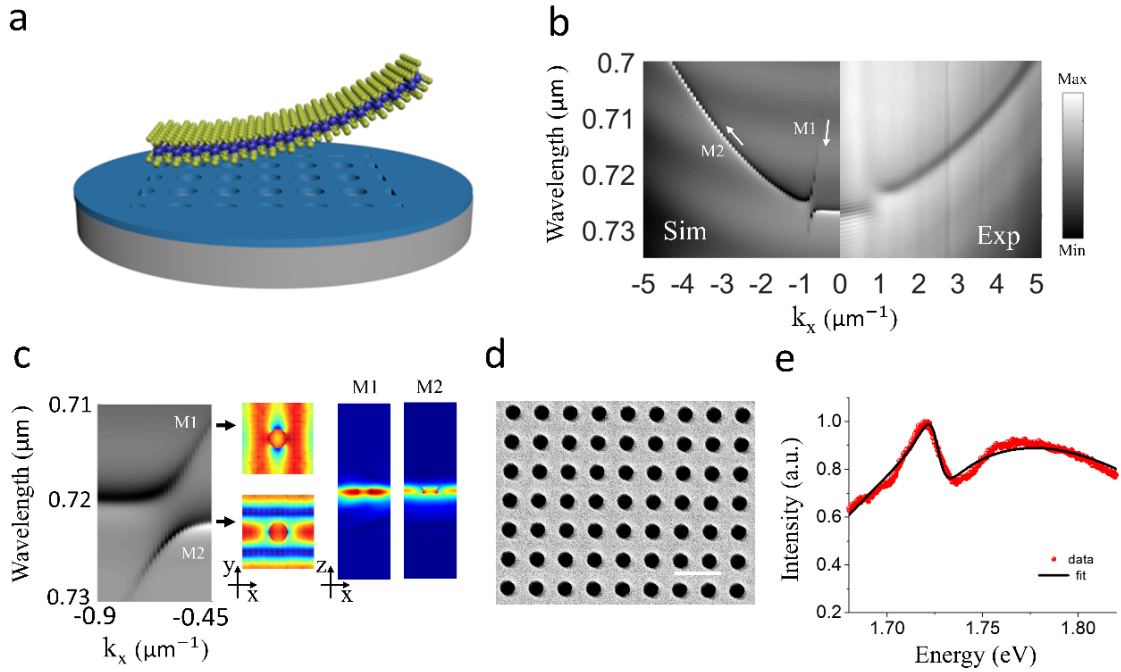

*Figure 1: SiN metasurface supporting guided mode resonances: (a) The metasurface is made of SiN meta-atoms with holes arranged in a square lattice. The whole structure sits on a silicon dioxide substrate. A WSe$_2$ monolayer could be transferred directly on top of the metasurface for evanescent coupling. (b) Simulated vs Experimentally measured angle-dependent reflection spectrum. There are two modes (M1 and M2) in the spectrum. M1 has a linear dispersion and starts at higher energy and rapidly goes to lower energy when it comes to high momentum. M2 has a parabolic shape and it starts at lower energy and goes to the higher energy. The two modes come close to each other at $k_x = 0.6$ μm$^{-1}$: an anti-crossing appears due to the coupling between the two photonics modes. (c) Zoom-in of the anti-crossing and the mode profiles of M1 and M2. Two modes have different symmetry in terms of the field distribution, but they are both well confined inside the metasurface. (d) The SEM of the SiN metasurface (scale bar: 1 μm) (e) Example spectrum and fitting at $k_x = 2.2$ μm$^{-1}$. A Fano-lineshape is observed with a Q factor ~143 and asymmetry factor -1.4.*

We then fabricated the metasurface and performed energy-momentum spectroscopy at room temperature to measure the angle-dependent reflectivity spectrum (optical setup is shown in the supplementary materials). The SEM image of the fabricated device is shown in the Figure 1d. The white light source is sent through a beam splitter and focused onto the sample by the objective lens (NA = 0.6). The image of the reflective light at the back focal plane (BFP) of the objective lens is then sent into the spectrometer. The slit at the entrance of the spectrometer (Princeton Instruments PIXIS CCD with an IsoPlane SCT-320 Imaging Spectrograph) acts as a spatial filter that only passed the signal with $k_y = 0\ \mu m^{-1}$. The grating inside the spectrometer then dispersed the signal ($k_x$, $k_y = 0\ \mu m^{-1}$) to the two-dimensional CCD sensor. As shown in the right panel of Figure 1b, the dispersion of the metasurface along the ($k_x$, 0) direction is detected through the CCD, where the x-axis is the $k_x$ wavevector and the y-axis is the wavelength. The result matches well with the simulation at the left side in Figure 1b. The mode-splitting due to the photonic-photonic mode coupling is also observed at $k_x = 0.6\ \mu m^{-1}$.

We then fit the reflection spectrum $R(\omega)$ at each $k_x$ value through a functional form[35]:

$$R(\omega) = R_{Fano} + R_{FP} + R_b \qquad (1)$$

Here, $R_{Fano}$ is the Fano linshape resulting from the interference of the GMR and the SiN slab mode:

$$R_{Fano} = I_0\left(1 - \frac{(x+q)^2}{x^2+1}\right) \qquad (2)$$

$$x = \frac{E-E_0}{\Delta\omega/2} \qquad (3)$$

Here, $I_0$ is the amplitude coefficient of the Fano resonance, $q$ is the asymmetry factor, $E = \hbar\omega$ is the photon energy at an angular frequency of $\omega$, and $E_0 = \hbar\omega_o$ is the photon energy at the cavity resonance frequency $\omega_o$, $\Delta\omega$ is the Full width at half maximum (FWHM). Since both $E_0$ and $\Delta\omega$ are in units of eV, $x$ becomes a unitless quantity. $R_{FP}$ is a broad Fabry-Perot (FP) interference background resulting from the reflection between the SiN/SiO$_2$ interface and the SiO$_2$/Si interface. We calculate $R_{FP}$ through the transfer matrix method[35] for the multiple thin film structure (SiN/SiO$_2$/Si). $R_b$ represents an ambient background in the fitting. Figure 1e shows an example of fitting for the reflection spectrum at $k_x = 2.2 \ \mu m^{-1}$. Through the fitting, the resonance energy $E_0$ is extracted as 1.726 eV and the Q factor of the resonance as ~143 ($\Delta\omega$ = 12 meV). The asymmetric q-factor is fit as q = -1.3.

**Exciton-polaritons in hybrid WSe$_2$-metasurface structure**

After characterizing the bare metasurface, we transferred a hBN encapsulated WSe$_2$ monolayer on top following usual dry transfer process[39]. The hBN encapsulation has been shown to improve the surface smoothness of the WSe$_2$ and preserve the intrinsic narrow linewidth of the WSe$_2$[48]. We then performed the energy-momentum spectroscopy on the hybrid WSe$_2$ on metasurface structure at 22K. As shown in Figure 2b, the GMR dispersion changed dramatically when coupled with WSe$_2$ at the exciton wavelength (~715nm). A clear anti-crossing is observed in the range from $k_x$ = 1.5 $\mu m^{-1}$ to $k_x$ = 3 $\mu m^{-1}$. We fit the dispersion spectrum at each $k_x$ value with the Fano-lineshape function to extract the resonance wavelengths of the upper and lower polaritons. An example fit spectrum at $k_x = 2.4 \ \mu m^{-1}$ is shown in Figure 2c, where two Fano lineshapes appear, corresponding to the upper polariton (UP) and lower polariton (LP). We note that, this spectrum is dramatically different from Figure 1e where only one Fano resonance is measured as the cavity resonance. For the spectrum shown in Figure 2c (at $k_x = 2.4 \ \mu m^{-1}$), the resonance energy for the UP and LP is found as 1.718 eV and 1.74 eV, respectively from the fitting parameters. We also observed the anti-crossing feature with another sample at 80K (Figure S2 in the supplementary materials).

Once the spectrum for each k-vector is fit and the wavelengths of the UPs and LPs are extracted, we estimate the loss and the interaction strength of the coupled system. We use a dispersive coupled-oscillator model to fit the wavelength of the LPB and UPB[12].

$$\begin{pmatrix} E_{exc} + i\gamma_{exc} & g \\ g & E_{cav} + i\gamma_{cav} \end{pmatrix} \begin{pmatrix} \alpha \\ \beta \end{pmatrix} = E_p \begin{pmatrix} \alpha \\ \beta \end{pmatrix} \quad (4)$$

Here $E_{exc}$ is the energy of the bare exciton and $\gamma_{exc}$ is its decay rate. $E_{cavity}$ is the energy of the bare cavity and $\gamma_{cav}$ is the cavity decay rate. g is the coupling strength between the exciton and the cavity. $E_p$ represents the eigenvalues corresponding to the energies of the polariton modes, and it could be found as:

$$E_{LP,UP} = \frac{1}{2}[E_{exc} + E_{cav} + i(\gamma_{cav} + \gamma_{exc})] \pm \sqrt{g^2 + \frac{1}{4}[E_{exc} - E_{cav} + i(\gamma_{exc} - \gamma_{cav})]^2} \quad (5)$$

α and β construct the eigenvectors and represent the weighting coefficients of the cavity photon and exciton for each polariton state, where $|\alpha|^2 + |\beta|^2 = 1$. The Hopfield coefficients which indicate the exciton and photon fraction in each LP and UP are given by the amplitude squared of the coefficients of eigenvectors ($|\alpha|^2$ and $|\beta|^2$).

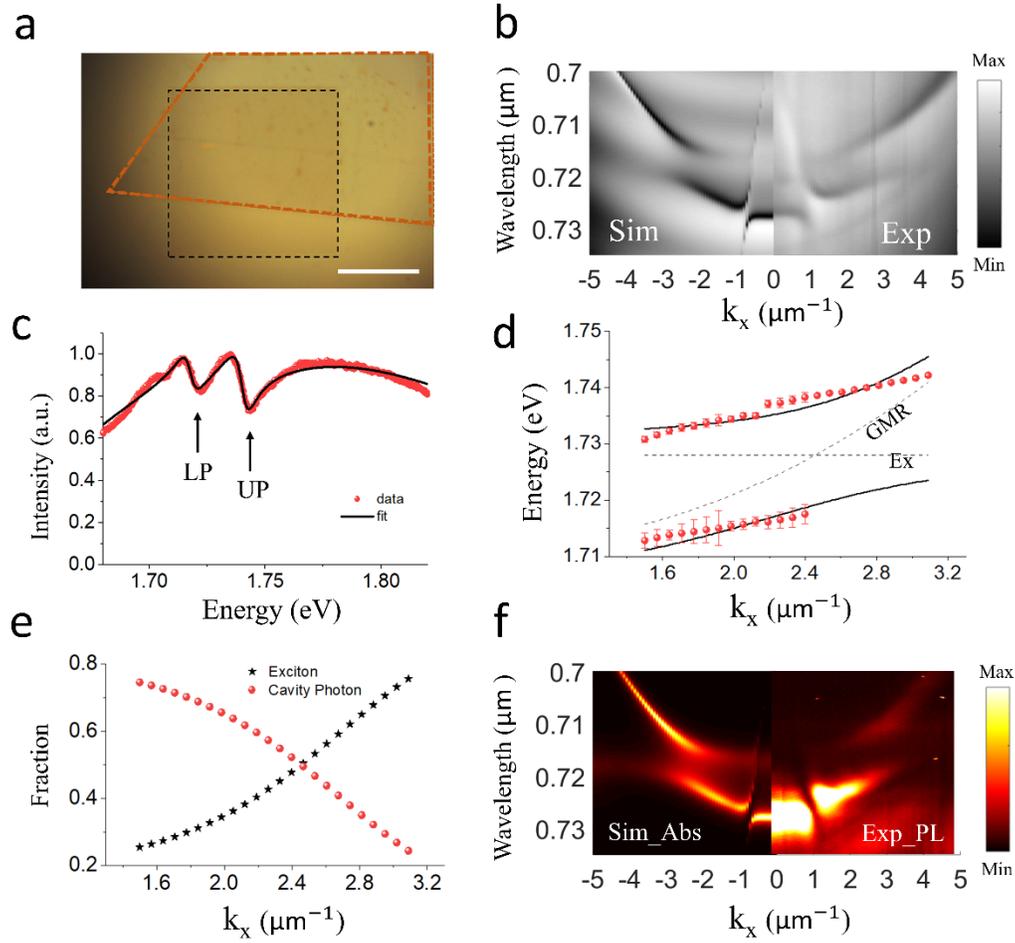

Figure 2 (a) The optical microscope image of the SiN metasurface with hBN-capped WSe$_2$ transferred (scale bar: 20 μm). The black lines outline the metasurface and the orange lines outline the monolayer WSe$_2$. The hBN is hardly observed under the microscope due to poor optical interference. (b) Simulated vs experimentally measured angle-dependent reflection spectrum. Anti-crossing is observed at $k_x \sim 2.4\ \mu m^{-1}$. (c) Example of a fit reflection spectrum at $k_x = 2.4\ \mu m^{-1}$, the Fano-resonance of the lower polariton (LP) and upper polariton (UP) are observed. (d) Fitting for the anti-crossing: a Rabi splitting value of the 18 meV is extracted. (e) Hopfield coefficients of the LP branch, which show the exciton and photon fraction in the polariton. (f) PL emission also shows the anti-crossing: LPB emission is brighter than UPB due to the thermal equilibrium condition.

Since we measured $E_{cavity}$ before the transfer of WSe$_2$ and $E_{LP,UP}$ after the transfer, the independent parameters in our fit are $\gamma_{exc}$, $E_{exc}$, g and $\gamma_{cav}$. During fitting, we shifted the cavity resonance ($E_{cavity}$) to account for the effect of the hBN and temperature-dependence of the cavity resonance. Through the fitting, we extracted $E_{exc} = 1.728 eV$, $\gamma_{exc} = 6.1\ meV$, $\gamma_{cav} = 8.3\ meV$ and g $= 9.1\ meV$. We also calculate the Rabi splitting $\hbar\Omega_{Rabi} = 2\sqrt{g^2 - \frac{1}{4}(\gamma_{exc} - \gamma_{cav})^2} = 18\ meV$. This value is of the similar magnitude of the values reported using one-dimensional photonic lattice[35,49]. We theoretically estimate the Rabi splitting to be $\sim 66\ meV$ (details in the supplementary materials). The ~3 times reduction of the measured Rabi splitting can be attributed to unwanted air gap between the WSe$_2$ monolayer and metasurfaces due to wrinkles in the monolayer or residues from the material transfer.

We then compared the dissipation rate with the interaction strength to confirm that we are indeed in the strong-coupling regime. The conditions to reach strong coupling are:

$$g > |\gamma_{exc} - \gamma_{cav}|/2 \text{ and } g > \sqrt{(\gamma_{exc}^2 + \gamma_{cav}^2)/2} \qquad (6)$$

The first is to guarantee a non-vanishing Rabi splitting, the second is to guarantee that the Rabi splitting is larger than the exciton and cavity linewidth so that the splitting could be experimentally measured. Clearly, the extracted parameters ($\gamma_{exc} = 6.1\ meV$, $\gamma_{cav} = 8.3\ meV$ and g $= 9.1\ meV$) satisfy above two criteria. We also measured photoluminescence (PL) and observed the anti-crossing in PL (Figure 2b). The PL is brighter at LPB than UPB as expected from the thermal equilibrium condition[50]. We also calculated the angle-dependent Hopfield coefficient ($\alpha$ and $\beta$) from the coupled-oscillator model (Eqn. 4). Figure 2e shows the Hopfield coefficients of the LPB. The LPB is more photon-like for $k_x < 2.4\ \mu m^{-1}$ and more excitonic-like for $k_x > 2.4\ \mu m^{-1}$.

Finally, we validate the experiment results with the numerical simulation. In RCWA simulation, we added a monolayer of WSe$_2$ (thickness = 0.7nm) on top of the SiN meta-atom to simulate the reflection/absorption spectrum of our WSe$_2$-metasurface structure. The dielectric function of the monolayer of WSe$_2$ is described by a Lorentz model:

$$\varepsilon(E) = \varepsilon_B + \frac{f}{E_x^2 - E^2 - i\Gamma E} \tag{7}$$

The $\varepsilon_B$ is the background dielectric constant, $f$ is the oscillator strength, $\Gamma$ is the linewidth of the exciton, and $E$ ($E_x$) represents the photon (exciton resonance) energy. For WSe$_2$, the value of $\varepsilon_B$ is 25, and we use the value of $E_x$ and $\Gamma$ from our previous fitting using Equation 5 ($E_x$ = 1.728 eV and $\Gamma = 2 \times \gamma_{exc}$ = 12.2 meV). We used oscillator strength f = 0.7 $eV^2$ to reproduce the Rabi splitting observed in the experiment[35]. As shown in the left panels of Figure 2b and Figure 2f, our simulated reflection and absorption spectra show good agreement with the experimental reflection and PL spectra (right panels of Figure 2b and Figure 2f).

The nano-patterned subwavelength structure can also produce a highly directional polariton emission in the far-field. Here we focus on the emission from the lower polariton since it is significantly brighter than the upper polariton. A 720 nm short-pass filter is placed at the collection path to collect the signal only from the lower polariton. Figure 3a shows the back focal plane image of the spatial distribution of the lower polariton far-field emission. Different from the emission pattern of an in-plane exciton dipole on an unpatterned substrate[51], the polariton emission shows a unique pattern: the metasurface efficiently diffracts the emission into (k$_x$, k$_y$) = (0,0) $\mu m^{-1}$ direction. The maximum values of the k$_x$ and k$_y$ axis are governed by the numerical aperture of the objective lens. In fact, the maximum k value collected at the back focal plane of the lens is given by $k_{max} = \frac{2\pi NA}{\lambda}$, where $\lambda$ is the wavelength of the lower polariton emission (~730 nm) and NA is the numerical aperture of the lens (in our case NA = 0.6). Figure 3b shows the emission in the far-field along the ($k_x = 0$ $\mu m^{-1}$, $k_y$) direction: a peak around $k_x = 0$ is clearly observed. Through a Lorentzian fitting we extracted the FWHM = $0.13 k_{max}$. The FWHM could then be converted to divergence angle through the relation $\theta = 2\,arcsin\left(\frac{1}{2} \times \frac{FWHM}{2\pi/\lambda}\right)$. The final divergence angle is calculated to be *5º*. The unique spatial dispersion of the polariton in far-field would have potential applications in the development of compact 2D monolayer polariton laser with unidirectional emission property.

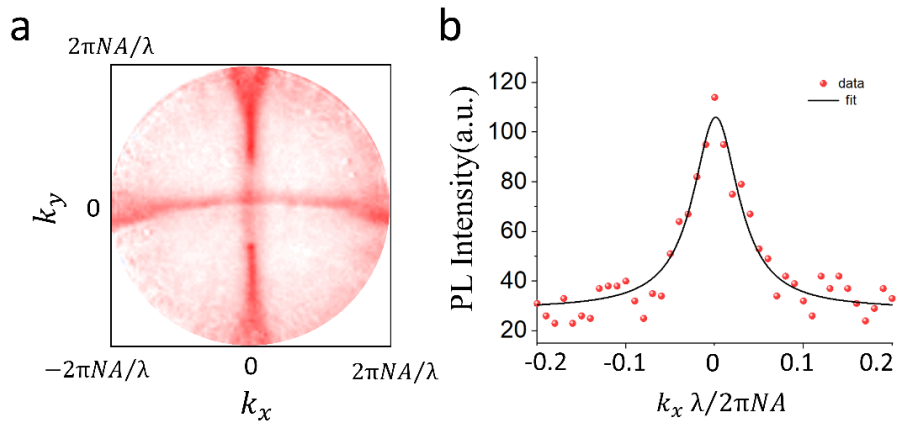

*Figure 3 (a) Back focal plane (BFP) image of the lower polariton (LP) far-field emission. Different from the excitonic emission of a monolayer on a unpattern substrate, the polariton emission shows a unique pattern due to the diffraction effect of the nanophotonic structures. (b) The emission in the far-field along the ($k_x$ = 0, $k_y$) direction, an emission peak around $k_x$ = 0 is observed. By fitting using a Lorentzian function, we extracted the divergence angle of the unidirectional emission on the order of 5º.*

## Discussion

**Meta-optical engineering of the exciton-polariton**

One unique property of our metasurface-based exciton-polariton system is the ability to engineer the Rabi splitting, the polariton dispersion and the far-field emission pattern by exploiting the large number of degrees of freedom offered by the nanopatterned photonic structures. Here we systematically study how the properties of the exciton-polariton platform can be tailored by engineering the optical meta-atoms, including their thickness, duty cycle, periodicity and lattice type.

We first study the effect of the metasurface thickness to the light-matter interaction strength. Different from the traditional DBR-sandwiched monolayer platform, the $WSe_2$ is evanescently coupled with the metasurface. As a result, the light-matter coupling strength (g) is proportional to the electrical field at the surface of the meta-atoms. This effect could be quantified by normalizing the electrical field on each meta-atom to its vacuum energy as[52,53]:

$$\int_{meta-atom} \varepsilon \psi^2 dV = \hbar\omega \qquad (8)$$

Here, $\varepsilon$ is the dielectric function of the meta-atom and the $\psi$ is the electric field. The $\hbar\omega$ represents the vacuum energy of the GMR supported by the meta-atom. This equation indicates that a well-confined mode leads to a large normalized electric field amplitude. However, the surface field also suffers from a stronger exponential decay in the out-of-plane direction as the confinement becomes stronger[54]. Here we define 'surface field' as the electric field at the surface of the meta-atom with which the $WSe_2$ monolayer would interact (equation S1 in the supplementary materials). As a result, there is a trade-off between the amplitude of the normalized electric field and the amount of surface field. We simulated the eigenmodes in the meta-atom with different thickness using COMSOL Multiphysics and normalized the electric field according to the equation (8). Then we calculated the amplitude of the surface field (Figure 4a). The surface field amplitude slightly increases as the thickness increases from 65nm to 100nm, and then decreases with the further increase of the slab thickness. The results indicate a slab thickness ~100nm would be the optimal thickness for enhancing light-matter coupling in our platform. To further validate our result, we simulated the Rabi splitting of the $WSe_2$ coupled with the meta-atom with various thicknesses via RCWA simulation. The period of the meta-atom is adjusted while the thickness is changed to match the photonic resonance to the $WSe_2$ exciton wavelength (715nm). As shown in Figure 4a, the trend of the Rabi splitting qualitatively matches with the trend of the normalized surface electric field.

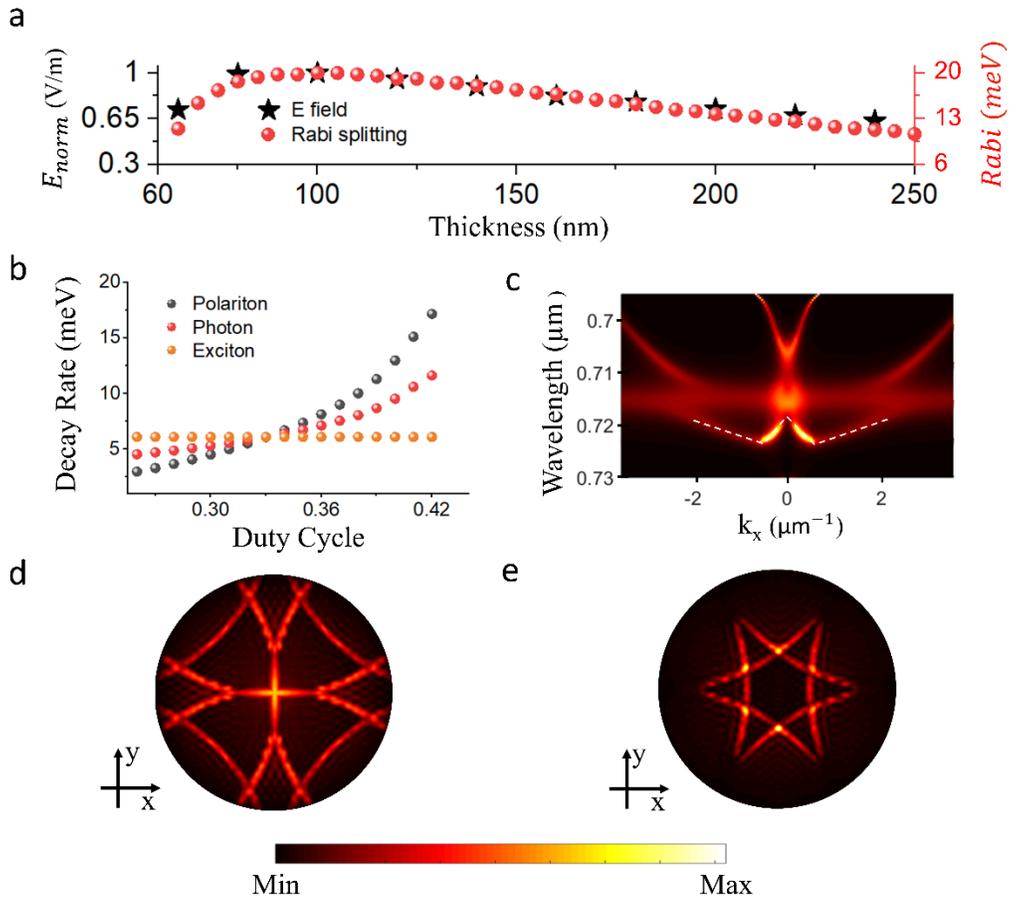

*Figure 4 (a) Slab thickness-dependent Rabi splitting and normalized E field. The results indicate a slab thickness ~100nm would be the optimal thickness for enhancing the light-matter coupling strength in our platform. (b) The effect of the duty cycle of the hole in the meta-atom to the polariton decay rate. A small duty ratio would effectively suppress the polariton decay rate into the free space. (c) A unique W-shape polariton dispersion in our platform (The white dash line). (d)(e) Polariton far-field emission pattern for square and hexagonal lattice, respectively.*

We then explored the effect of the duty cycles of the hole in the meta-atom to the polariton decay rate. The duty cycle is defined as the ratio of the hole diameter to the lattice periodicity. We simulated the photon decay rate as a function of the duty cycle with a fixed period. As shown in the Figure 4b, the decay rate increases when the hole size becomes larger. Using Equation 5 and experimentally extracted exciton decay rate, we calculated the polariton decay rate when the photon wavelength matches with the exciton resonance. The result in the Figure 4b indicates that a small duty cycle would effectively suppress the polariton decay rate. Moreover, in the strong coupling regime, the excited-state lifetime could be strongly modified by controlling the coupling of the cavity mode to the radiation continuum, which can be achieved via subwavelength engineering of the meta-atoms. Recent theoretical study predicts that by coupling $WSe_2$ to a bound state-in-continuum supported by a periodic photonic lattice, it is possible to achieve a 100-fold enhancement of the polariton lifetime by suppressing the coupling to the radiation continuum[55].

We also analyzed the effect of the polarization and the periodicity in our exciton-polariton system. Different from a one-dimensional photonic lattice, the s and p polarizations are degenerate at normal incident ($k_x = 0$) due to the intrinsic symmetry of our metasurface. This degeneracy is lifted with gradually increasing k value. Such degeneracy would allow the study of valley-polariton[18]. Figure S3 in the supplementary material shows the reflection spectrum of the s and p polarization of the metasurface before integrating with the monolayer. A rich cluster of dispersion behavior, including linear, parabolic and W-shape dispersion, is supported in the momentum-space of the exciton-polariton system. The various slopes of dispersion favor future study of our polariton platform both in the high and slow group-velocity regime, with application in ballistic propagation of polaritons[56] and Bose−Einstein condensation[57]. By properly adjusting the period of the meta-atom, a photonic mode with specific shape could be strongly coupled to the exciton. We performed an RCWA simulation where we adapted the metasurface parameters that we used to simulate for the Figure S3, in addition, adding a monolayer of $WSe_2$ on top of the structure. Figure 4c showed the absorption spectrum of the hybrid structure, where the coherent coupling of the exciton to a

'W-shaped' dispersion is observed (the white dash line). This unique W-shaped dispersion would favor the future study on exotic polariton physics, for example, momentum-space Josephson effect[58] and micro-optical parametric oscillation[59].

Finally, we investigated the effect of the photonic lattice type in the far-field emission pattern. We first simulate the diffraction pattern for a rectangular lattice (details in the supplementary materials). As shown in Figure 4d, the far-field emission has two different orders of diffraction. We were able to resolve the first order of diffraction in our experiment as shown in Figure 3a. The higher-order diffraction could not be resolved possibly due to the extra loss and limited NA of our setup. With the fixed hole size and period, we then slightly change the lattice into hexagonal one. Surprisingly, the far-field emission pattern dramatically changes into a star-shape (Figure 4e). The ability to tailor the far-field emission pattern would pave the way for the future development of polariton light-emitting devices.

In conclusion, we demonstrated exciton-polaritons in atomically thin tungsten diselenide ($WSe_2$) strongly coupled to the GMR in a SiN metasurface. The strong coupling regime is probed via energy-momentum spectroscopy on the $WSe_2$-metasurface sample, and a coherent light-matter interaction strength of ~18meV is measured. Finally, we showed that the Rabi splitting, the polariton dispersion and the far-field emission pattern could be tailored by subwavelength-scale engineering of the optical meta-atoms in the metasurface. Our platform opens the door for the future development of novel exciton-polariton devices by advanced meta-optical engineering.

# Method

**Metasurface fabrication:** The SiN metasurface is fabricated on a silicon nitride on oxide chip that consists of a 130nm of PECVD SiN, 1 μm of oxide and 500 μm silicon substrate. We spun coat ZEP520A resist on it and patterned using a JEOL JBX6300FX with an accelerating voltage of 100kV. The pattern was transferred to the silicon nitride using an inductive-coupled plasma etching in CHF3/SF6 chemistry. The ZEP520A resist is latter stripped off by organic solution.

# Acknowledgement

This work is supported by the National Science Foundation under grant NSF MRSEC 1719797. S. Miao, T. Wang and S.-F. Shi acknowledge support from AFOSR (Grant No. FA9550-18-1-0312), ACS PRF (Grant No. 59957-DNI10) and a KIP award from RPI. Part of this work was conducted at the Washington Nanofabrication Facility / Molecular Analysis Facility, a National Nanotechnology Coordinated Infrastructure (NNCI) site at the University of Washington, which is supported in part by funds from the National Science Foundation (awards NNCI-1542101, 1337840 and 0335765), the National Institutes of Health, the Molecular Engineering & Sciences Institute, the Clean Energy Institute, the Washington Research Foundation, the M. J. Murdock Charitable Trust, Altatech, ClassOne Technology, GCE Market, Google and SPTS.

# Supplementary Information: Metasurface Integrated Monolayer Exciton Polariton

Yueyang Chen[1]†, Shengnan Miao[4]†, Tianmeng Wang[4]†, Ding Zhong[2], Abhi Saxena[1], Colin Chow[2], James Whitehead[1], Xiaodong Xu[2,3], Sufei Shi[4,5], Arka Majumdar[1,2]

[1] Electrical and Computer Engineering, University of Washington, Seattle, WA 98189, USA
[2] Department of Physics, University of Washington, Seattle, WA 98189, USA
[3] Materials Science and Engineering, University of Washington, Seattle, WA 98189, USA
[4] Department of Chemical and Biological Engineering, Rensselaer Polytechnic Institute, Troy, New York 12180, USA
[5] Department of Electrical, Computer, and Systems Engineering, Rensselaer Polytechnic Institute, Troy, New York 12180, USA

† These authors contributed equally to this work.

## S1. Optical setup for energy-momentum spectroscopy

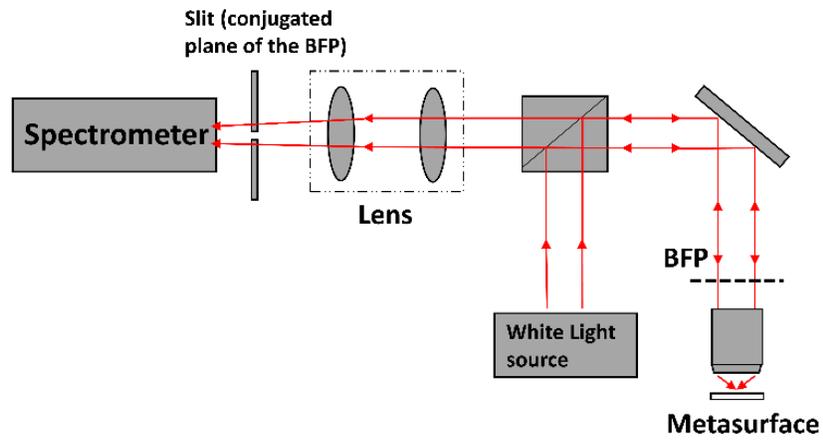

*Figure S1: Optical setup for energy-momentum spectroscopy. The back focal plane image was focused to the slit of the spectrometer by a lens with focal length of 150 mm followed with a telescope setup to optimize the size of the image.*

## S2. Anti-crossing in PL at 80K

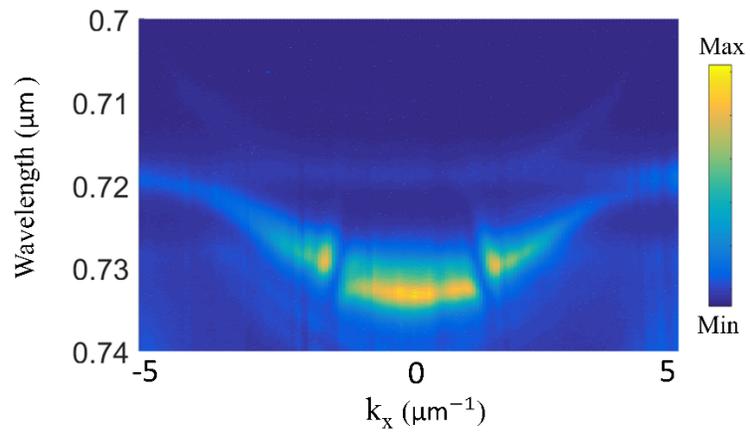

*Figure S2: PL data at 80K with another sample. An anti-crossing is observed ~720nm.*

## S3. Metasurface reflection for s and p polarization

To study the polarization-dependent reflection, we perform RCWA simulation on the meta-atom with slab thickness of 130 nm, period of 463 nm and duty ratio of 0.5.

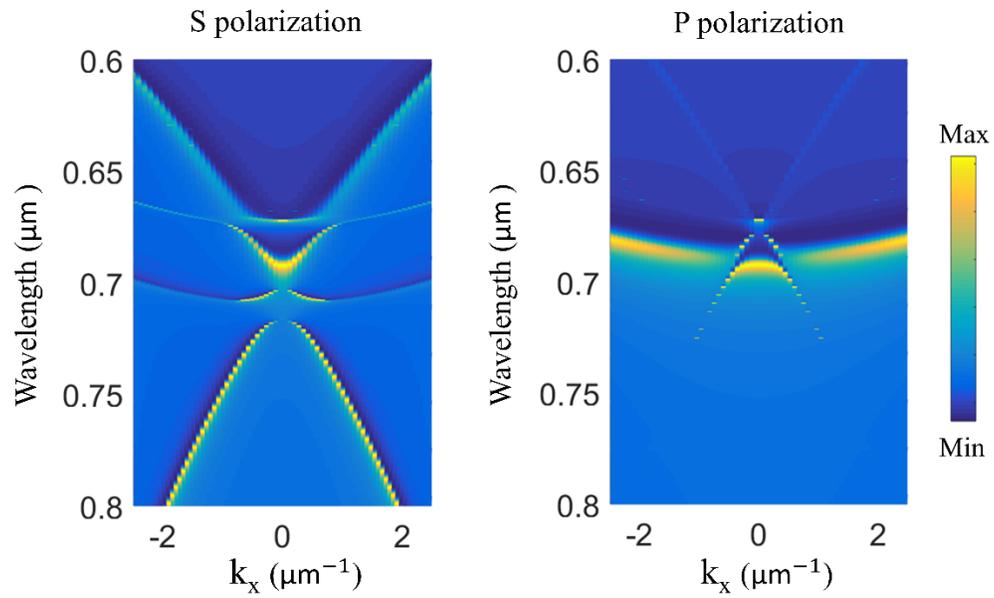

*Figure S3: s and p polarization response in reflection obtained via RCWA simulation*

## S4. Calculation of the surface electric field of the meta-atom

We calculated the surface electric field of the mate-atom through COMSOL Multiphysics. In the Finite Element simulation, the x and y boundary are set as Bloch boundary condition while the z boundaries are set with PML boundary. We picked a fixed value of $(k_x, k_y) = (2, 0)$ $\mu m^{-1}$ to solve for the eigenmode, while changing the thickness of the slab from 65nm to 250nm. We traced for the mode 2 in Figure 2c of main text. As thickness is increased, the wavelength of the eigen-mode slightly red-shifted, which we compensate by changing the dimension of the meta-atom. We then normalized the electrical field by the vacuum energy of the eigenmode (Equation 8 of the main text). Form the normalized electrical field, we calculated the average surface field as:

$$\psi_{avg} = \frac{1}{a^2} \int_{Surface} |\psi| \, dS \tag{S1}$$

where $a$ is the period of the meta-atom. The results are shown in the Figure 4a.

## S5. Far-field simulation

We calculated the far-field emission via Lumerical FDTD simulation to study the diffraction effect of the metasurface. Due to the computational limitation, we simulate an area covering $40 \times 40$ unit cells and we put two incoherent electrical dipoles with s and p polarization to simulate the emission from the monolayer. We record the electromagnetic field in the near-field monitor slightly above the metasurface. And then we calculated the far-field using the build-in far-field projector. The results are shown in the Figure 4d in main text, where the simulation matches well with the experimental data in Figure 3a (in main text).

## S6. Calculation of the vacuum Rabi splitting

We theoretically estimated the Rabi splitting of our coupled WSe$_2$-Metasurface system as[1]:

$$\hbar\Omega_{Rabi} = 2\hbar\eta\sqrt{\frac{2\Gamma_0 c}{n_c L}}$$

Here, $\hbar$ is the plank constant and c is the speed of light. $\Gamma_0 = 1/2\tau$ is the exciton radiative broadening, where $\tau$ is the intrinsic radiative lifetime of the WSe$_2$ exciton. We used the value of intrinsic radiative lifetime as the theoretically calculated value ($\tau = 0.22$ ps)[2]. $n_c$ represents the refractive index of the cavity, here we use the refractive index of silicon nitride ($n_c = 2$). L represent the lateral thickness of the cavity, which is equal to 130nm in the case of our metasurface. $\eta$ is defined as $\frac{\int_{Surface}|\psi|\,dS}{\int_{Maximum}|\psi|\,dS}$ since the monolayer is evanescently coupled to the optical mode in the metasurface. We calculated $\eta = 0.7$ from COMSOL Multiphysics. Using the above parameters, we calculated the Rabi Splitting $\hbar\Omega_{Rabi} = 66$ meV.